\documentclass[10pt,twocolumn,twoside]{IEEEtran}
\usepackage{amsfonts,dsfont,amssymb,amsbsy,amsmath,paralist,theorem,bm,ifthen,color}
\usepackage[pdfstartview=FitH,bookmarksnumbered,unicode,bookmarksopen=true]{hyperref}
\usepackage{graphicx}
\usepackage{epstopdf}
\usepackage{multirow}
\usepackage{subfigure,relsize}
\usepackage{mathrsfs}
\usepackage{bm}
\usepackage{stfloats}
\usepackage{url}
\usepackage[noadjust]{cite}
\usepackage{cases,booktabs}
\usepackage{graphicx,algorithmic,algorithm,relsize}
\usepackage[justification=centering]{caption} 
\usepackage[table]{xcolor}

\newtheorem{lem}{Lemma}
\newtheorem{prop}{Proposition}

\newtheorem{coro}{Corollary}
\newtheorem{rem}{Remark}

\newcommand\pin{\ensuremath{{\rm Pin}}}

\newcommand\wb{\ensuremath{{\bf w}}}

\newcommand\ub{\ensuremath{{\bm u}}}

\newcommand\hb{\ensuremath{{\bf h}}}

\newcommand\Xb{\ensuremath{{\bf X}}}
\newcommand\xb{\ensuremath{{\bf x}}}

\newcommand\Vb{\ensuremath{{\bf V}}}
\newcommand\vb{\ensuremath{{\bf v}}}

\newcommand\psib{\ensuremath{{\bm \psi}}}

\newcommand\wo{\ensuremath{{\textrm{wo}}}}
\newcommand\w{\ensuremath{{\textrm{w}}}}

\def\CN{{{\mathcal{CN}}}}

\newcommand\E{\ensuremath{{\mathbb{E}}}}

\newcommand\SNR{\ensuremath{{\rm SNR}}}

\newcommand\Cs{\ensuremath{{\mathbb{C}}}}

\newcommand\Nset  {\ensuremath{{\mathcal{N}}}}
\newcommand\Mset  {\ensuremath{{\mathcal{M}}}}

\newcommand\st    {\ensuremath{{\rm s.t.}}}

\newcommand\Hf{\ensuremath{{\mathsf{H}}}}

\graphicspath{{fig/}}

\definecolor{green}{RGB}{34	195	46}
\definecolor{red}{RGB}{220 0 0}

\usepackage{graphicx} 

\title{Pinching-Antenna Systems with In-Waveguide Attenuation: Performance Analysis and \\Algorithm Design}

\author{
 Yanqing Xu, \IEEEmembership{Member, IEEE,}
         Zhiguo Ding, \IEEEmembership{Fellow, IEEE,}
         Robert Schober, \IEEEmembership{Fellow, IEEE,}
         and Tsung-Hui Chang, \IEEEmembership{Fellow, IEEE}
         \thanks{\smaller[1] Y. Xu is with the School of Science and Engineering, The Chinese University of Hong Kong, Shenzhen, 518172, China (email: xuyanqing@cuhk.edu.cn).}
         \thanks{\smaller[1] Z. Ding is with the University of Manchester, Manchester, M1 9BB, UK (email: zhiguo.ding@manchester.ac.uk).}
         \thanks{\smaller[1] R. Schober is with the Institute for Digital Communications, Friedrich-Alexander-University Erlangen-N\"{u}rnberg (FAU), 91054 Erlangen, Germany (e-mail: robert.schober@fau.de).}
         \thanks{\smaller[1] T.-H. Chang is with the School of Artificial Intelligence, The Chinese University of Hong Kong, Shenzhen, 518172, China (email: changtsunghui@cuhk.edu.cn).} 
         \thanks{This work has been submitted to the IEEE for possible publication. Copyright may be transferred without notice, after which this version may no longer be accessible.}
        }

\date{\today}

\begin{document}

\maketitle

\begin{abstract}
    Pinching-antenna systems have emerged as a promising flexible-antenna architecture for next-generation wireless networks, enabling enhanced adaptability and user-centric connectivity through antenna repositioning along waveguides. However, existing studies often overlook in-waveguide signal attenuation and in the literature, there is no comprehensive analysis on whether and under what conditions such an assumption is justified. This paper addresses this gap by explicitly incorporating in-waveguide attenuation into both the system model and algorithm design, and studying its impact on the downlink user data rates. We begin with a single-user scenario and derive a closed-form expression for the globally optimal antenna placement, which reveals how the attenuation coefficient and the user-to-waveguide distance jointly affect the optimal antenna position. Based on this analytical solution, we further provide a theoretical analysis identifying the system conditions under which the in-waveguide attenuation has an insignificant impact on the user achievable rate. The study is then extended to the multi-user multiple-input multiple-output setting, where two efficient algorithms are developed—based on the weighted minimum mean square error  method and the maximum ratio combining method—to jointly optimize beamforming and antenna placement. Simulation results validate the efficacy of the proposed algorithms and demonstrate that pinching-antenna systems substantially outperform conventional fixed-antenna baselines, underscoring their potential for future flexible wireless communications.
\end{abstract}


\section{Introduction} 
The next generation of wireless communication systems is expected to meet increasingly demanding requirements in terms of system throughput, latency, connectivity density, and energy efficiency. To support emerging applications, such as immersive extended reality, holographic communications, autonomous driving, and the Internet of Everything (IoE), future networks must have the capability to deliver gigabit-per-second data rates with ultra-low latency and high reliability across diverse deployment environments \cite{you2021towards}. Among the enabling technologies, multi-antenna systems play a pivotal role due to their ability to enhance spectral efficiency, spatial diversity, and user multiplexing \cite{xu2025distributed,larsson2014massive}.
However, existing multi-antenna solutions are predominantly based on fixed-position antenna systems, where antenna elements are pre-deployed at static positions, such as base station (BS) towers or distributed access points. These fixed deployments are often suboptimal due to spatial channel variations, dynamic user mobility, and limited coverage flexibility, especially in scenarios with complex or rapidly changing environments. Therefore, new architectural innovations are needed to fully exploit the spatial degrees of freedom offered by multi-antenna systems. Motivated by the aforementioned issues, flexible-antenna systems have been recently developed to make both the positions of the antennas as well as wireless channels of the users adjustable \cite{wong2020fluid,zhu2023movable}. However, existing flexible-antenna systems still lack array reconfiguration flexibility and are incapable of mitigating line-of-sight (LoS) blockage or reducing large-scale path loss.

Pinching-antenna system has recently emerged as a novel and practical flexible-antenna system to overcome the aforementioned limitations of conventional antenna systems \cite{suzuki2022pinching,ding2024flexible}. In this system, antennas—referred to as pinches—are mechanically mounted on a waveguide and can be dynamically moved to or activated at arbitrary positions on the waveguide tailored to the served users' positions. This mechanical flexibility brings several key advantages. First, by allowing each pinching antenna to be repositioned closer to its served user, strong LoS channels can be established even in challenging propagation environments. Second, the antenna system is inherently flexible: new pinching antennas can be added or removed with minimal modifications to the infrastructure, enabling scalable deployment and adaptation to user density. Third, from a practical standpoint, pinching-antenna systems offer a cost-effective alternative to traditional distributed antenna systems, such as cloud radio access networks, distributed multiple-input multiple-output (MIMO), or cell-free MIMO. In particular, those architectures rely on spatially distributed remote radio heads and hence are constrained by fronthaul capacity and synchronization requirements, whereas pinching-antenna systems avoid such issues by maintaining centralized processing and physical connectivity via the waveguides. Notably, pinching antennas are not a replacement but rather a complementary enhancement to other distributed antenna systems, and can be jointly implemented within the same network infrastructure.

Several recent studies have explored the great potential of pinching-antenna systems from various perspectives \cite{ding2024flexible,yang2025pinching,liu2025pinching,zhou2025channel,xiao2025channel,xu2025rate,tegos2025minimum,tyrovolas2025performance,zhou2025gradient,wang2025antenna,xu2025qos,fu2025power,zhou2025sum,xie2025graph,xu2025joint,wang2025modeling,bereyhi2025mimo,zhang2025two,ding2025blockage,qin2025joint,ding2025pinching,khalili2025pinching,ouyang2025rate,mao2025multi,jiang2025pinching}. 
The work in \cite{ding2024flexible} systematically demonstrated the ability of pinching antennas to dynamically adjust antenna positions and enhance user-centric service delivery. 
For channel estimation, both model-driven \cite{zhou2025channel} and learning-based \cite{xiao2025channel} approaches have been proposed to exploit the structured propagation characteristics of pinching-antenna systems. 
From a data rate maximization perspective, pinching-antenna position optimization for downlink and uplink pinching-antenna systems were investigated in \cite{xu2025rate} and \cite{tegos2025minimum}, respectively. The outage probability of pinching-antenna system was studied in \cite{tyrovolas2025performance} by taking in-waveguide attenuation into consideration.
Leveraging the characteristics that only one data stream can be passed through a waveguide, non-orthogonal multiple access (NOMA) assisted pinching-antenna systems were investigated in \cite{wang2025antenna,xu2025qos,fu2025power,zhou2025sum}. In such schemes, the signals for multiple users are superposed based on NOMA and then fed into a waveguide for information delivery. 
Pinching antennas have also been used for MIMO communications \cite{wang2025modeling,bereyhi2025mimo,zhang2025two}, where multiple pinching antennas are deployed on multiple waveguides to serve multiple downlink users. By jointly optimizing the downlink beamfomer and pinching-antenna positions, the spectral and energy efficiencies are significantly improved. More recently, pinching antennas have also been applied in other communication scenarios. 
For example, several works have explored the integration of pinching antennas with integrated sensing and communication (ISAC) systems \cite{ding2025pinching,khalili2025pinching,qin2025joint,ouyang2025rate,mao2025multi}. Specifically, the work \cite{ding2025pinching} leveraged the Cram\'{e}r–Rao lower bound framework to demonstrate how pinching antennas can ensure uniform positioning accuracy across users, thanks to the flexible deployment of pinching antennas. In a complementary direction, the authors of \cite{khalili2025pinching} proposed dynamically activating pinching antennas to exploit look-angle diversity, thereby enhancing target sensing reliability in ISAC networks. Additionally, the authors of \cite{jiang2025pinching} examined how pinching antennas can be leveraged for covert communications by exploiting dynamic antenna placement to reduce detectability.

Despite the promising advancements in prior studies on pinching-antenna systems, a common limitation lies in the oversight of in-waveguide attenuation, with little  theoretical analysis of its impact on system performance.
To bridge this gap, we explicitly incorporate in-waveguide attenuation into both system modeling and algorithm design, aiming to maximize downlink data rates of users.
This approach enables a more realistic and comprehensive understanding of the design principles of pinching-antenna systems.
The key contributions of this work are summarized as follows:
\begin{itemize}
    \item We first focus on the single-user scenario. In this case, we derive the globally optimal closed-form solution for the antenna position that maximizes the user’s data rate in the presence of in-waveguide attenuation.
    This solution reveals that the optimal pinching-antenna position depends on both the in-waveguide attenuation coefficient and the user-to-waveguide distance.
    Moreover, based on this solution, we provide a rigorous theoretical analysis demonstrating that the performance degradation caused by overlooking the in-waveguide attenuation becomes negligibly small if system parameters, including waveguide height, attenuation coefficient, and communication region size, satisfy certain mild and practically realizable conditions. 
    This result not only quantifies the conditions under which attenuation can be safely ignored but also offers concrete design guidelines for deploying pinching-antenna systems.

    \item We then extend the study to the more general multi-user MIMO (MU-MIMO) scenario. In this case, the joint optimization of beamforming vectors and pinching-antenna positions becomes significantly more challenging due to the strong coupling between the channel coefficients and the antenna positions. To address this, we first reformulate the problem using the classical weighted minimum mean square error (WMMSE) framework \cite{shi2011iteratively}, which enables iterative updates of the transmit beamformers and pinching-antenna positions. However, since the channel coefficients depend on the antenna positions in a highly non-convex manner, the antenna position updates must be performed via a linear search, which can become computationally expensive as the number of antennas increases. To balance performance and computational efficiency, we further propose a two-stage algorithm that first employs a low-complexity maximum ratio combining (MRC)-based approximation to determine the antenna positions, and then applies the WMMSE method to optimize the beamforming vectors given the obtained antenna positions.
    
    \item Extensive numerical results are provided to demonstrate the effectiveness of  pinching-antenna systems and the efficacy of the proposed algorithms. The results show that the proposed two-stage WMMSE-MRC algorithm achieves a performance comparable to that of WMMSE-based method, while significantly reducing computational complexity. In addition, the pinching-antenna system consistently outperforms the conventional fixed-position antenna system for various scenarios, including different transmit power levels, coverage areas, and user densities, demonstrating its robustness and potential for practical deployment in future flexible wireless networks.
    
\end{itemize}

This paper is organized as follows: Section~\ref{sec: siso} presents the system model and formulates the data rate maximization problem for a single-user pinching-antenna system with in-waveguide attenuation. A closed-form solution for the optimal antenna placement is derived, and a theoretical analysis is conducted to characterize the data rate loss when in-waveguide attenuation is ignored. 
Section~\ref{sec: mu-mimo} extends the study to the MU-MIMO setting. A WMMSE-based algorithm is proposed to jointly optimize beamforming and antenna positions, and a low-complexity two-stage WMMSE-MRC algorithm is developed to improve computational efficiency.
Section~\ref{sec: simulation} validates the  efficacy of the proposed methods and theoretical findings through extensive simulations for various system parameters.
Finally, Section~\ref{sec: conclusion} concludes the paper.

{\bf Notations:} Column vectors and matrices are denoted by boldfaced lowercase and uppercase letters, e.g., $\xb$ and $\Xb$.
$\mathbb{C}^{N \times 1}$ stand for the sets of $N$-dimensional complex vectors.
The superscripts $(\cdot)^\top$ and $(\cdot)^\Hf$ describe the transpose and Hermitian operations, respectively.
$||\xb||^2$ denotes the square of the Euclidean norm of vector ${\xb}$. $\mathcal{R}\{C\}$ returns the real part of a complex-valued number $C$.
$\E_n[\cdot]$ represents the statistical expectation operation with respect to random variable $n$.

\section{Pinching-Antenna System Design with In-Waveguide Attenuation: The Single-User Case} \label{sec: siso}
\begin{figure}[!t]
	\centering
	\includegraphics[width=0.96\linewidth]{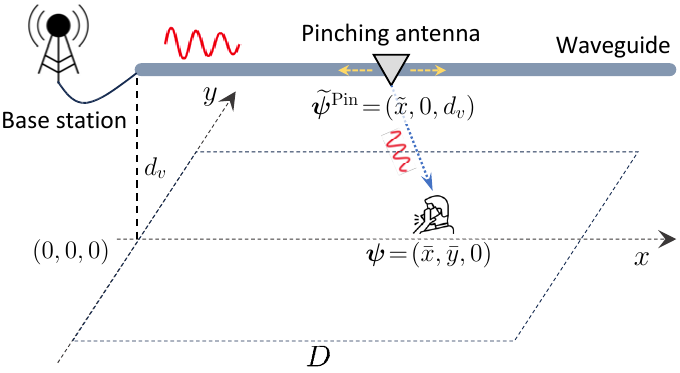}\\
        \captionsetup{justification=justified, singlelinecheck=false, font=small}	
        \caption{The considered pinching-antenna systems with a single user.} \label{fig: system model siso} 
\end{figure}

In this section, the case with a single user is considered. Specifically, we first focus on the single-input single-output (SISO) case, where a single pinching antenna is activated on a waveguide and serves a single-antenna user, as shown in Fig. \ref{fig: system model siso}. Without loss of generality, we assume that the user is uniformly distributed over a square communication region with side length $D$, and its location is denoted by $\psib = [\bar x, \bar y,0]$, where $0 \leq \bar x \leq D$ and $-\frac{D}{2} \leq \bar y \leq \frac{D}{2}$.
The waveguide is deployed at the center of this region, parallel to the $x$-axis, with a fixed height $d_v$. The coordinate of the feed point of the waveguide is $\bm{\psi}_0 = [0,0,d_v]$. 
The position of the pinching antenna is denoted by $\widetilde\psib^{\pin} = [\tilde x, 0, d_v]$, respectively.

\subsection{Channel Model with In-Waveguide Attenuation}
We assume that there is a LoS link between the user and the pinching antenna, so the free-space channel model is adopted \cite{ding2024flexible}.
Then, the channel coefficient between the feed point and the user is given by
\begin{small}
\begin{align}\label{eqn: channel model siso}
    &h = \underbrace{\frac{\eta^{\frac{1}{2}}}{\|\bm{\psi} - \boldsymbol{\widetilde{\psi}}^{\pin}\|}}_{\text{free-space path loss}} 
     \underbrace{e^{-\alpha \|\bm{\psi}_0 - \boldsymbol{\widetilde{\psi}}^{\pin}\|}}_{\text{waveguide attenuation}} 
     \underbrace{e^{-j \left( \frac{2\pi}{\lambda} \|\bm{\psi} - \boldsymbol{\widetilde{\psi}}^{\pin}\| + \frac{2\pi}{\lambda_g} \|\bm{\psi}_0 - \boldsymbol{\widetilde{\psi}}^{\pin}\| \right)}}_{\text{dual phase shift}}.
\end{align}
\end{small}%
As can be seen, the channel coefficient is determined by three components:
\begin{itemize}
    \item The first term characterizes the \textit{free-space path loss} between the $n$-th pinching antenna and the user, where $\eta = \frac{c}{4\pi f_c}$ is a constant with $c$ denoting the speed of light and $f_c$ representing the carrier frequency.
    \item The second term models the \textit{in-waveguide attenuation}, which causes an exponential power loss due to the signal propagation within the waveguide over distance $\|\bm{\psi}_0 - \boldsymbol{\widetilde{\psi}}^{\pin}\|$ \cite{pozar2021microwave}. Here, $\alpha$ denotes the attenuation coefficient\footnote{The waveguide attenuation coefficient depends on several factors, including the waveguide material, structure, and operating frequency. For instance, in a typical dielectric waveguide, the value of the waveguide attenuation coefficient is approximately $0.08$ dB/m \cite{ji2017ultra,bauters2011planar}.}. A higher $\alpha$ results in a greater power loss for longer distances within the waveguide.
    \item The third term captures the \textit{dual phase shift} due to signal propagation in both free space and within the waveguide. Specifically, $\frac{2\pi}{\lambda}\|\bm{\psi} - \boldsymbol{\widetilde \psi}^{\pin}\|$ is the phase shift caused by signal propagation in free space, and $\frac{2\pi}{\lambda_g}\|\bm{\psi}_0 - \boldsymbol{\widetilde \psi}^{\pin}\|$ denotes the phase shift due to signal propagation inside the waveguide. Here, $\lambda$ is the free-space wavelength and $\lambda_g = \frac{\lambda}{n_{\rm neff}}$ denotes the guided wavelength with $n_{\rm neff}$ denoting the effective refractive index of the dielectric waveguide, which typically ranges between $1.0$ and $2.0$ depending on the waveguide material and geometry \cite{pozar2021microwave}.
\end{itemize}
By inserting the respective coordinates into the channel model in \eqref{eqn: channel model siso}, the channel coefficient can be expressed as
\begin{align}\label{eqn: channel coefficient siso}
    h &= \frac{\eta^{\frac{1}{2}} e^{-j \big( \frac{2\pi}{\lambda} \big[(\tilde x - \bar x)^2 + C\big]^{\frac{1}{2}} + \frac{2\pi}{\lambda_g} \tilde x \big)} }{\big[(\tilde x - \bar x)^2 + C\big]^{\frac{1}{2}} e^{\alpha\tilde x}},
\end{align}
where $C = \bar y^2 + d_v^2$ is the square of the distance between the user and the closest point on the waveguide.

Compared to the channel model of conventional flexible-antenna systems, the proposed model in \eqref{eqn: channel coefficient siso} exhibits several unique characteristics that highlight the distinct nature of pinching-antenna systems. First, the ability to place the pinching antenna at any position along the waveguide provides increased flexibility in adjusting the large-scale path loss, as the antenna positions can be flexibly tuned to shorten propagation distances and enhance LoS conditions. Second, unlike traditional systems where channel phase shift is typically induced by free-space propagation alone, the pinching-antenna channel exhibits two phase shifts: one arising from the external free-space transmission and the other from signal propagation inside the waveguide. In other words, the phase of a channel in pinching-antenna systems can also be reconfigured by adjusting the position of the pinching antenna. Third, the model incorporates in-waveguide attenuation, an additional factor absent in conventional pinching-antenna models, which can affect the amplitude of the received signal and must be carefully managed. These combined features introduce additional degrees of freedom for improving channel quality and enabling user-centric adaptation. However, they also bring about new research challenges for practical algorithm design.

\subsection{Problem Formulation}
In this work, we are interested in maximizing the downlink transmission data rate of the pinching-antenna system. The associated problem can be formulated as follows
\begin{subequations}
    \begin{align}
        \max_{\tilde x} ~&\log \Big(1 + \frac{P|h|^2}{\sigma^2} \Big) \\
        \st~& 0 \leq \tilde x \leq x_{\max}, \label{eqn: location constraint siso}
    \end{align}
\end{subequations}
where $P$ is the maximum transmission power of the pinching antenna, and $\sigma^2$ is the received additive white Gaussian noise (AWGN) power at the user. Constraint \eqref{eqn: location constraint siso} characterizes the feasible region for the position of the pinching antenna.

Note that maximizing the data rate is equivalent maximizing the received signal-to-noise ratio (SNR).  The associated problem is given by
\begin{align} \label{p: siso}
    \max_{0 \leq \tilde x \leq x_{\max}}~ \frac{\rho \eta}{(\tilde x - \bar x)^2 e^{2\alpha\tilde x} + C e^{2\alpha\tilde x}},
\end{align}
where $\rho = P/\sigma^2$. The denominator of \eqref{p: siso} captures the combined effects of waveguide attenuation and free-space path loss on the effective channel gain and, consequently, on the received SNR. Specifically, as the pinching antenna moves away from the feed point of the waveguide, the waveguide attenuation increases exponentially, while the free-space path loss decreases. This creates an interesting trade-off between waveguide attenuation and free-space path loss based on the pinching-antenna position. 

\subsection{Derivation and Analysis of the Optimal Solution of Problem \eqref{p: siso}} \label{sec: solution siso}
In this subsection, we first derive the optimal solution to problem \eqref{p: siso}. Next, we provide an in-depth analysis of how the system parameters, specifically, the waveguide attenuation parameter $\alpha$ and the square of the distance between the user and the waveguide $C$, affect the optimal pinching-antenna position. The optimal position of the pinching antenna is established in the following lemma.

\begin{lem}\label{lem: siso} 
    The optimal pinching-antenna position is given by
    \begin{align}\label{eqn: optimal location siso}
        \tilde x^* = 
        \begin{cases}
            0, & \textrm{if~} C \geq \frac{1}{4\alpha^2 } - \frac{(2\alpha \bar x - 1)^2}{4\alpha^2 }, \\
            \bar{x} + \frac{-1 + \sqrt{1 - 4\alpha^2 C}}{2\alpha}, & \textrm{otherwise}.
        \end{cases}
    \end{align} 
\end{lem}

{\textit{Proof:}} First, it is not difficult to verify that the optimal pinching-antenna position satisfies $x^* \leq \bar x$. We notice that problem \eqref{p: siso} is equivalent to finding that $x \in [0, \bar x]$ which minimizes
\begin{align}
    f(\tilde x) = (\tilde x - \bar x)^2 e^{2\alpha\tilde x} + C e^{2\alpha\tilde x}. 
\end{align}
To find the point minimizing $f(\tilde x)$, we need to check its first-order derivative, which is given by
\begin{align}
    f'(\tilde{x}) = e^{2\alpha \tilde{x}} \left[ 2 (\tilde{x} - \bar{x}) + 2\alpha (\tilde{x} - \bar{x})^2 + 2\alpha C \right].
\end{align}
Since $e^{2\alpha \tilde{x}} > 0$, we focus on the following function:
\begin{align}
    g(\tilde{x}) \triangleq \alpha (\tilde{x} - \bar{x})^2 + (\tilde{x} - \bar{x}) + \alpha C.
\end{align}
First, let's rewrite the quadratic function $g(\tilde{x})$ as
\begin{align}
    g(\tilde{x}) = \alpha \Big(\tilde{x} - \bar{x} + \frac{1}{2\alpha}\Big)^2 + \alpha C - \frac{1}{4 \alpha}.
\end{align}
We note that if $C \geq \frac{1}{4\alpha^2 }$, $g(\tilde{x}) \geq 0$ always holds. In this case, $f(\tilde x)$ is monotonically increasing, and thus the optimal pinching-antenna position is $\tilde x^* = 0$ for this case. 

For the case that $C < \frac{1}{4\alpha^2 }$, the quadratic equation $g(\tilde{x}) = 0$ has two real roots, which are given by
\begin{align}
    x_1 &= \bar{x} + \frac{-1 - \sqrt{1 - 4\alpha^2 C}}{2\alpha},\\
    x_2 &= \bar{x} + \frac{-1 + \sqrt{1 - 4\alpha^2 C}}{2\alpha}, 
\end{align}
where $x_1 < x_2 < \bar x$ holds. 
Note that if $x_2 \leq 0$, i.e., $C \geq \frac{1}{4\alpha^2 } - \frac{(2\alpha \bar x - 1)^2}{4\alpha^2 }$, we have that $g(\tilde{x}) > 0$ for $x \in [0, \bar x]$. In this case, $f(\tilde x)$ is a monotonically increasing function of $\tilde x \in [0, \bar x]$. Therefore, the optimal pinching-antenna position is $ \tilde x^* = 0$.

For the case that $x_2 \geq 0$, i.e., $C \leq \frac{1}{4\alpha^2 } - \frac{(2\alpha \bar x - 1)^2}{4\alpha^2 }$, we have that $|2\alpha \bar x -1| < \sqrt{1 - 4\alpha^2 C}$. Then, we can verify that $x_1 < 0$. This indicates that $g(\tilde{x}) < 0$ for $x \in [0, x_2)$ and $g(\tilde{x}) \geq 0$ for $x \in [x_2,\bar x)$.
Therefore, $f(\tilde x)$ is monotonically decreasing for $x \in [0, x_2)$ and monotonically increasing for $x \in [x_2, \bar x)$.
Thus, the optimal pinching-antenna position is $\tilde x^* = x_2$. Summarizing the above steps, the proof of Lemma \ref{lem: siso} is complete. \hfill $\blacksquare$

From the proof of Lemma \ref{lem: siso}, we can draw several interesting observations:
\begin{enumerate}
    \item When $\alpha$ is small, i.e., the waveguide attenuation is not significant, such that $C < \frac{1}{4\alpha^2 } - \frac{(2\alpha \bar x - 1)^2}{4\alpha^2 } = \frac{\bar x}{\alpha} - \bar x^2$, the optimal pinching-antenna position is $x_2$. This indicates that the pinching-antenna position should be properly selected to balance the waveguide attenuation and the free-space path loss to maximize the effective channel gain.
    
    \item When $\alpha \rightarrow 0^+$, we have that $C < \frac{1}{4\alpha^2 } - \frac{(2\alpha \bar x - 1)^2}{4\alpha^2 }$ always holds, because
    \begin{align}
        \lim_{\alpha \rightarrow 0^+} \frac{1}{4\alpha^2 } - \frac{(2\alpha \bar x - 1)^2}{4\alpha^2 } = \lim_{\alpha \rightarrow 0^+} \frac{\bar x}{\alpha} - \bar x = + \infty.
    \end{align}
    Then, the optimal pinching-antenna position is $x_2$, and we have
    \begin{align}
        &\lim_{\alpha \rightarrow 0^+} \bar{x} + \frac{-1 + \sqrt{1 - 4\alpha^2 C}}{2\alpha} \notag\\
        \overset{(a)}{=}& \lim_{\alpha \rightarrow 0^+} \bar{x} + \frac{- 2\alpha^2 C + O(\alpha^4)}{2\alpha} \notag\\
        =& \lim_{\alpha \rightarrow 0^+} \bar{x} - \alpha C + O(\alpha^3) = \bar x, \label{eqn: low attenuation}
    \end{align}
    where, in $(a)$, the Taylor series of the term $\sqrt{1 - 4\alpha^2 C}$ is applied with $\mathcal{O}(\alpha^4)$ denoting the higher-order terms that are asymptotically bounded by a term propotional to $\alpha^4$ as $\alpha \rightarrow 0$ \cite{rudin1964principles}. The result in \eqref{eqn: low attenuation} indicates that when the waveguide attenuation is negligible, it suffices to deploy the pinching antenna at $[\bar x, 0, d_v]$, as this position offers the smallest free-space path loss. 

    \item When $C$ is small, such that $C < \frac{1}{4\alpha^2 } - \frac{(2\alpha \bar x - 1)^2}{4\alpha^2 } = \frac{\bar x}{\alpha} - \bar x^2$, the optimal pinching-antenna position is $x_2$. Meanwhile, as $C$ decreases, the term $-1 + \sqrt{1 - 4\alpha^2 C}$ approaches $0$, and thus $x_2$ gradually converges to $\bar x$. This indicates that  when the user is near the waveguide, the pinching antenna should be deployed close to the user, making the impact of in-waveguide attenuation on antenna placement less significant.

    \item When $C$ is large (for a given $\alpha$), such that $C \geq \frac{1}{4\alpha^2 }$, the free-space path loss dominates the overall path loss. In this case, the relative benefit of adjusting the antenna position along the waveguide becomes negligible, and Lemma \ref{lem: siso} suggests to place the pinching antenna directly at the feed point to avoid unnecessary waveguide attenuation.
    
    \item When $\alpha$ is large (for a given $C$), such that $C \geq \frac{1}{4\alpha^2 }$, even small shifts along the waveguide introduce substantial signal attenuation due to the high in-waveguide loss factor. To mitigate this, the antenna should again be placed at the feed point.


    \item When $\alpha$ and $C$ are moderate such that $C \leq \frac{1}{4\alpha^2} - \frac{(2\alpha \bar{x} - 1)^2}{4\alpha^2}$, Lemma \ref{lem: siso} indicates the following behavior of the effective channel gain:
    \begin{itemize}
        \item For $0 \leq \tilde{x} \leq x_2$, the effective channel gain is more sensitive to the free-space path loss;
        \item For $x_2 \leq \tilde{x} < \bar{x}$, the effective channel gain is primarily dominated by waveguide attenuation, owing to the reason that the waveguide attenuation increases exponentially with the distance between the feed point of the waveguide and the pinching antenna position.
    \end{itemize}

\end{enumerate}

\begin{figure}[!t]
	\centering
	\includegraphics[width=0.96\linewidth]{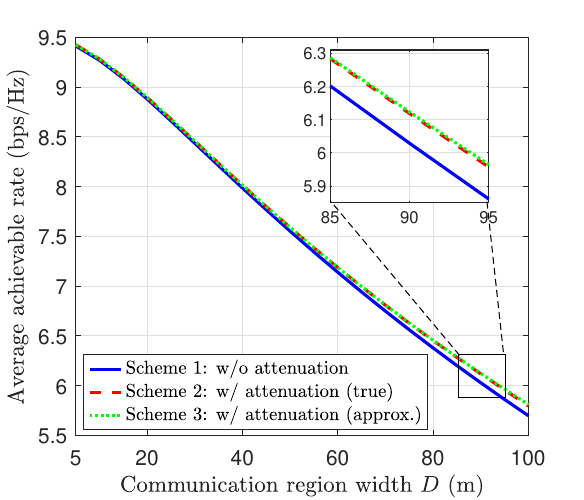}\\
        \captionsetup{justification=justified, singlelinecheck=false, font=small}	
        \caption{Average achievable rate versus communication region width $D$ for three design schemes in a SISO pinching-antenna system.} \label{fig:rate_vs_D}  
\end{figure} 

Beyond the intuitive insights into how the system parameters influence pinching antenna placement, it is also important to further examine their impact on system performance. To this end, we numerically compare the average achievable rate versus communication region width $D$ in Fig. \ref{fig:rate_vs_D}, where the system parameter are set as $d_v=10$ m, $\alpha = 0.0092$ m$^{-1}$ (corresponds to $\alpha = 0.08$ dB/m in decibel units), $f = 28$ GHz, $P = 10$ Watts and $\rho = -70$ dBm. All three considered schemes take in-waveguide attenuation into account when evaluating the system performance (i.e., computing the achievable data rate). However, they differ in how the pinching-antenna position is determined. In particular, Scheme $1$ determines the antenna position under the idealized assumption of negligible in-waveguide attenuation, placing the antenna on the position on the waveguide that is closest to the user (i.e., $\tilde x = \bar x$); Scheme $2$ determines the antenna position by explicitly considering the in-waveguide attenuation, which given by \eqref{eqn: optimal location siso}; Scheme $3$ approximates the optimal antenna position in \eqref{eqn: optimal location siso} by directly setting $\tilde x^* = \bar{x} + \frac{-1 + \sqrt{1 - 4\alpha^2 C}}{2\alpha}$, which corresponds to the solution assuming a sufficiently long waveguide.

As observed, the performance of all considered schemes degrades as $D$ increases due to longer average user distances. The performance gap between Scheme $1$ and Scheme $2$ demonstrates the rate loss caused by ignoring in-waveguide attenuation during design, which is modest but grows with the region width $D$. This gap is also numerically illustrated in Fig. \ref{fig:rate_gap_vs_D}
Notably, Scheme $3$ closely tracks Scheme $2$, confirming that the approximate antenna placement is roughly accurate, and thus serves as a reliable upperbound approximation of the optimal scheme.

\subsection{Average Data Rate Analysis and System Design Insights} \label{subsec:ergodic_SNR}
To gain deeper analytical insights and derive quantitative design guidelines, this subsection presents a theoretical analysis of the average achievable data rate under different pinching antenna design strategies. Specifically, we focus on comparing the averaged achievable rates of the two schemes with and without considering in-waveguide attenuation. Notably, Scheme $3$ has been shown through simulations to closely approximate the optimal design. Therefore, we adopt it as the reference scheme for capturing the effects of in-waveguide attenuation in the theoretical analysis.

\begin{figure}[!t]
	\centering
	\includegraphics[width=0.96\linewidth]{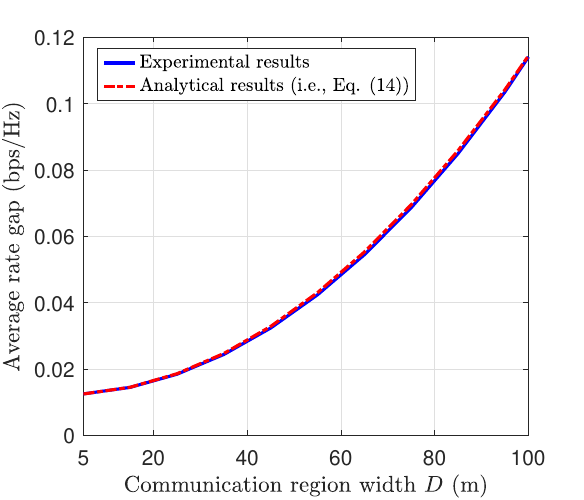}\\
        \captionsetup{justification=justified, singlelinecheck=false, font=small}	
        \caption{Comparison between the analytical expression in \eqref{eqn: rate gap} and simulation results for the average rate gap versus communication region width $D$.} \label{fig:rate_gap_vs_D} 
\end{figure}

In this section, we are interested in quantifying the achievable data rate loss due to ignoring in-waveguide attenuation during pinching-antenna placement. Specifically, we evaluate how this loss varies under different system configurations, including communication region size $D$, waveguide height $d_v$, and in-waveguide attenuation coefficient $\alpha$. The following proposition provides a closed-form approximation of the average data rate loss incurred when the pinching antenna is positioned without accounting for in-waveguide attenuation.

\begin{prop} \label{prop: expected rate loss}
    The average data rate loss $\mathbb{E}_{\psib}[\Delta R]$ incurred by optimizing the antenna position without accounting for in-waveguide attenuation can be approximated by
    \begin{align} \label{eqn: rate gap}
        \mathbb{E}_{\psib}[\Delta R] \approx \frac{\alpha^2}{\ln 2} \left(\frac{D^2}{12} + d_v^2 \right) \ \ \ {{\textrm{(bps/Hz)}}},
    \end{align}
    where $\Delta R = R_{\w} - R_{\wo}$ is the difference in the achievable data rates with and without in-waveguide attenuation.
\end{prop}

\begin{IEEEproof}
    See Appendix \ref{appd: expected rate loss}.
\end{IEEEproof}

To validate the accuracy of the analytical result in \eqref{eqn: rate gap}, we compare it with simulation-based experimental results in Fig.~\ref{fig:rate_gap_vs_D}. As observed, the analytical expression closely matches the experimental results across all the considered communication region widths.
Based on Proposition \ref{prop: expected rate loss}, we can further derive a practical design rule that characterizes how large the communication region can be while keeping the expected rate loss below a predefined threshold. The following corollary establishes a relationship between the maximum allowable region size and a target rate loss $\epsilon$.
\begin{coro} \label{coro: rate loss D}
    To ensure that the average data rate loss due to neglecting in-waveguide attenuation is less than a predefined threshold $\epsilon > 0$ (bps/Hz), the communication region side length $D$ should satisfy:
    \begin{align}
        D \leq \sqrt{12 \left(\frac{\epsilon \ln 2}{\alpha^2} - d_v^2\right)}.
    \end{align}
\end{coro}

\begin{rem}
    Corollary \ref{coro: rate loss D} provides a quantitative guideline for practical system design. For a given attenuation level $\alpha$ and waveguide height $d_v$, one can compute the maximum allowable region width $D$ to ensure that the performance degradation caused by ignoring in-waveguide attenuation remains within acceptable bounds. For example, consider a system with waveguide height $d_v = 10$ m and attenuation coefficient $\alpha = 0.0092$ m$^{-1}$. To constrain the average data rate loss to no more than $\epsilon =0.1 $ bps/Hz, the maximum allowable communication region width $D$ can be calculated as:
    \begin{align}
        D \leq \sqrt{12 \left(\frac{0.1 \times \ln 2}{0.0092^2} - 10^2\right)} \approx \sqrt{8627} \approx 92.88 \ \textrm{m}.
    \end{align}
    This implies that to keep the average data rate loss below $0.1$ bps/Hz, the communication region width should not exceed approximately $92.88$ m under the given setting. 
\end{rem}

\subsection{Extension to the MISO Scenario} \label{sec: miso}
In this subsection, the MISO case is focused on, where $N$ pinching antennas deployed on $N$ waveguides are jointly used to serve a single-antenna user, as shown in Fig. \ref{fig: system model mu-mimo}. The waveguides are positioned parallel to the $x$-axis at a height $d_v$, and the distance between two neighboring waveguides is $d_h = \frac{D}{N-1} \gg \lambda$. 
Without loss of generality, we assume that the coordinate of the feed point of the $n$-th waveguide is $\bm{\psi}_{0,n} = [0,(n-1)d_h - \frac{D}{2},d_v]$. The position of the pinching antenna on the $n$-th waveguide is denoted by $\widetilde\psib_n^{\pin} = [\tilde{x}_n, (n-1) d_h - \frac{D}{2}, d_v]$. Then, the channel coefficient between pinching antenna $n$ and the user is given by
\begin{align}
    h_n = \frac{\eta^{\frac{1}{2}} e^{-j \big( \frac{2\pi}{\lambda} \big[(\tilde x_n - \bar x)^2 + C_n\big]^{\frac{1}{2}} + \frac{2\pi}{\lambda_g} \tilde x_n \big)} }{\big[(\tilde x_n - \bar x)^2 + C_n\big]^{\frac{1}{2}} e^{\alpha\tilde x_n}}, \forall n \in \Nset,
\end{align}
where $C_n = \big((n-1)d_h - \frac{D}{2} - \bar y\big)^2 + d_v^2$ is the square of the distance between the user and the $n$-th waveguide, and $\Nset \triangleq \{1,...,N\}$. 
The channel vector between the $N$ pinching antennas and the user is represented by $\hb = [h_{1},..., h_{N}]^\top$.
Let $\vb \in \Cs^{N \times 1}$ denote the beamformer for the user, the received signal at the user is given by
\begin{align} \label{eqn: received signal}
    y = \sqrt{P}\hb^\top \vb s + n,
\end{align}
where $s \in \Cs$ is the information symbol for the user, and $n \sim \CN(0,\sigma^2)$ denotes the AWGN at the user.  
Since there is only one user, MRC beamforming maximizes the received SNR at the user. In particular, the beamformer is given by $\vb = \frac{\hb}{\|\hb\|_2}$ and the received SNR at the user is given by $\SNR = \rho \|\hb\|^2$.
The design problem becomes:
\begin{subequations} \label{eqn: snr maximization miso}
    \begin{align}
        \max_{\tilde x_1,...,\tilde x_N}~& \sum_{n=1}^N  \frac{\rho \eta}{(\tilde x_n - \bar x)^2 e^{2\alpha\tilde x_n} + C_n e^{2\alpha\tilde x_n}}\\
        \st ~& 0 \leq \tilde x_n \leq x_{\max}, \forall n \in \Nset. \label{eqn: location constraint miso}
    \end{align}
\end{subequations}
Note that problem \eqref{eqn: snr maximization miso} a decomposable function of the positions of the pinching antennas, allowing us to optimize the each pinching-antenna positions separately. In particular, the problem for the $n$-th pinching-antenna position is given by
\begin{align} \label{p: subproblem miso}
    \max_{0 \leq \tilde x_n \leq x_{\max}}~ \frac{\rho \eta}{(\tilde x_n - \bar x)^2 e^{2\alpha\tilde x_n} + C_n e^{2\alpha\tilde x_n}},
\end{align}
which is identical to problem \eqref{p: siso} in the SISO case. Therefore, the optimal position for the $n$-th pinching antenna can be established in the following corollary:

\begin{figure}[!t]
	\centering
	\includegraphics[width=0.96\linewidth]{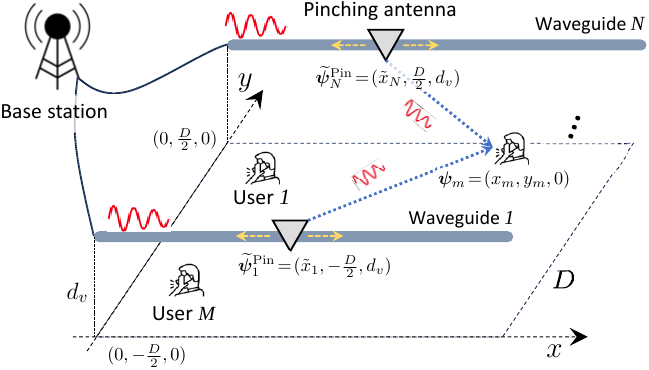}\\
        \captionsetup{justification=justified, singlelinecheck=false, font=small}	
        \caption{The considered downlink multiuser pinching-antenna systems.} \label{fig: system model mu-mimo} 
\end{figure} 

\begin{coro}
    The optimal position of the $n$-th pinching antenna is given by
    \begin{align} \label{eqn: pa location miso}
        \tilde x_n^* = 
        \begin{cases}
            0, & \textrm{if~} C_n \geq \frac{1}{4\alpha^2 } - \frac{(2\alpha \bar x - 1)^2}{4\alpha^2 },  \\
            \bar{x} + \frac{-1 + \sqrt{1 - 4\alpha^2 C_n}}{2\alpha}, & \textrm{otherwise}.
        \end{cases}
    \end{align}
\end{coro}

\section{Pinching-Antenna System Design with In-Waveguide Loss: The Multi-User Case} \label{sec: mu-mimo}
In this section, we consider a multi-user scenario where $N$ pinching antennas are deployed on $N$ waveguides and are jointly used to serve $M$ ($M \leq N$) single-antenna users, as shown in Fig. \ref{fig: system model mu-mimo}. 
The position of user $m$ is denoted by $\psib_m = [x_m, y_m, 0], \forall m \in \mathcal{M} \triangleq \{1,2,...,M\}$.
One viable approach in this case is that the users are served individually, for example using time division multiple access, where user $m$ is served in the $m$-th time slot. As a result, the solution proposed in Subsection \ref{sec: miso} is directly applicable. However, this scheme suffers from a performance loss due to its failure of fully exploiting the spatial degree-of-freedoms offered by multiple pinching antennas. In this subsection, we focus on a scheme where all the users are simultaneously served by the multiple pinching antennas.

To maintain generality, we continue to adopt the channel model accounting for in-waveguide attenuation, presented in \eqref{eqn: channel model siso}, although Proposition \ref{prop: expected rate loss} has shown that its impact is insignificant under mild conditions on the system parameters.
Based on the considered system setup, the channel coefficient between pinching antenna $n$ and user $m$ is given by
\begin{align} \label{eqn: channel mu}
    h_{m,n} \!=\! \frac{\eta^{\frac{1}{2}} e^{-j \big( \frac{2\pi}{\lambda} \big[(\tilde x_n - x_m)^2 + C_{m,n}\big]^{\frac{1}{2}} + \frac{2\pi}{\lambda_g} \tilde x_n \big)}} {\big[(\tilde x_n - x_m)^2 + C_{m,n}\big]^{\frac{1}{2}} e^{\alpha\tilde x_n}},  \forall n \!\in \! \Nset,  m \!\in\! \mathcal{M}.
\end{align}
Denote the channel vector between the $N$ pinching antennas and user $m$ as $\hb_m = [h_{m,1},..., h_{m,N}]^\top$, and let $\vb_m \in \Cs^{N \times 1}$ represent the beamformer for user $m$, which means that the received signal at user $m$ can be written as
\begin{align} \label{eqn: received signal}
    y_m = \hb_m^\top \vb_m s_m + \sum_{i \neq m} \hb_m^\top \vb_i s_i + n_m,
\end{align}
where $s_m \in \Cs$ is the information symbol for user $m$, and $n_m \sim \CN(0,\sigma_m^2)$ denotes the received AWGN at user $m$ with $\sigma_m^2$ denoting the noise power. 

The aim of the paper is to maximize the system sum rate by jointly optimizing the
beamformers and the pinching-antenna positions while satisfying a total transmit power constraint.
The associated problem can be formulated as
\begin{subequations} \label{p: sum rate maximization}
    \begin{align}
        \max_{\tilde \xb, \Vb} ~&\sum_{m=1}^M \log_2\left(1 + \frac{|\hb_m^\top \vb_m|^2}{\sum_{i \neq m}|\hb_m^\top \vb_i|^2 + \sigma_m^2}\right)\\
        \st~ & \sum_{m=1}^M \|\vb_{m}\|^2 \leq P_{\max},  \label{eqn: papc}\\
        & 0 \leq \tilde x_n \leq x_{\max}, \forall n \in \Nset,
    \end{align}
\end{subequations}
where $\tilde \xb = [\tilde x_1,...,\tilde x_N]^\top$ collects the positions of the pinching antennas, $\Vb = [\vb_1,...,\vb_M]$ contains the beamformers for the users. 
Constraint \eqref{eqn: papc} is the total power constraint.
Problem \eqref{p: sum rate maximization} is particularly challenging to solve for the following two reasons. First, the channel vectors are intricately dependent on the pinching-antenna positions, which are themselves optimization variables. Second, the channel and beamforming vectors are also intricately coupled within the objective function, further complicating the optimization process.
 
\subsection{A WMMSE-based Algorithm for Solving Problem \eqref{p: sum rate maximization}} \label{subsec: wmmse alg for joint design}

\subsubsection{Reformulation by Using WMMSE Method}
The WMMSE method can be used to reformulate the nonconvex sum-rate maximization problem to a more tractable weighted mean square error (MSE) minimization problem \cite{shi2011iteratively}. This motivates us to apply the WMMSE method to transform problem \eqref{p: sum rate maximization}. To proceed, based on the signal model in \eqref{eqn: received signal}, we first define $\hat s_m = u_m y_m$ as the estimated source signal at user $m$. Here, $u_m \in \Cs$ denotes the MMSE receiver. Then, the MSE of estimating $s_m$ can be written as 
\begin{align}
    e_m = |1 - u_m\hb_m^\top\vb_m|^2 + \sum_{i \neq m} |u_m \hb_m^\top\vb_i|^2 + \sigma_m^2 |u_m|^2.
\end{align}
Then, based on a similar idea as in \cite[Theorem 1]{shi2011iteratively}, one can show that the optimal solutions to $\tilde \xb$ and $\Vb$ for problem \eqref{p: sum rate maximization} are equivalent to the solution of the following problem
\begin{subequations} \label{p: wmmse problem}
    \begin{align}
        \min_{\tilde \xb,\Vb,\ub, \wb} ~&\sum_{m=1}^M w_m e_m - \log w_m \\
        \st~ & \sum_{m=1}^M |v_{m,n}|^2 \leq P_{\max}, \\
        & 0 \leq \tilde x_n \leq x_{\max}, \forall n \in \Nset,
    \end{align}
\end{subequations}
where $\ub = [u_1,...,u_M]^\top$ and $\wb = [w_1,...,w_M]$.
Problem \eqref{p: wmmse problem} is focused on in the remainder of the paper instead of \eqref{p: sum rate maximization}, due to the simpler structure of the objective function in \eqref{p: wmmse problem} with respect to the optimization variables. Specifically, the objective is convex with respect to beamforming vectors $\vb_m$, which facilitates the optimization process. Additionally, it is convex in the channel vectors $\hb_m$, though not necessarily in the antenna positions $\tilde x_n$. This contrasts with the objective function in \eqref{p: sum rate maximization}, where $\hb_m$ appears within fractional terms, making the optimization more challenging. 

Noting that the variables are not coupled in the constraints of problem \eqref{p: wmmse problem}, we can use a block coordinate descent (BCD)-based algorithm to solve it iteratively. In particular, in the $(t+1)$-th iteration, the variables are updated as 
\begin{subequations}
    \begin{align}
        \!\ub^{t+1} &= \arg \min_{\ub} \mathcal{F}\big(\tilde \xb^{t},\Vb^{t},\ub,\wb^{t}\big) \label{eqn: u problem}\\
        \!\wb^{t+1} &= \arg \min_{\wb}\mathcal{F}\big(\tilde \xb^{t},\Vb^{t},\ub^{t+1},\wb\big),\label{eqn: w problem}\\
        \!\Vb^{t+1} &= \arg \min_{\sum_{m=1}^M \|\vb_{m}\|^2 \leq P} \mathcal{F}\big(\tilde \xb^{t},\Vb,\ub^{t+1},\wb^{t+1}\big), \label{eqn: V problem}\\
        \!\xb^{t+1} &= \arg \min_{0 \leq \tilde x_n \leq x_{\max}, \forall n} \mathcal{F}\big(\tilde \xb,\Vb^{t+1},\ub^{t+1},\wb^{t+1}\big),\label{eqn: x problem}
    \end{align}
\end{subequations}
where $\mathcal{F}\big(\tilde \xb,\Vb,\ub,\wb\big) \triangleq \sum_{m=1}^M w_m e_m - \log w_m$, $\tilde \xb^{t}$, $\Vb^{t}$, $\ub^{t}$, and $\wb^{t}$ denote the solutions of $\tilde \xb$, $\Vb$, $\ub$, and $\wb$ in the $t$-th iteration, respectively. 
Notably, the updates of $\ub$ in \eqref{eqn: u problem} and $\wb$ in \eqref{eqn: w problem} have
closed-form solutions, which are given by 
\begin{subequations}
    \begin{align}
        u_m^{t+1} &= \frac{(\hb_m^t)^\top \vb_m^t}{\sum_{i=1}^M |(\hb_m^t)^\top \vb_i^t|^2 + \sigma_m^2}, \forall m, \label{eqn: update of u}\\
        w_m^{t+1} &= (e_m^{t+1})^{-1}, \forall m, \label{eqn: update of w}
    \end{align}
\end{subequations}
while problem \eqref{eqn: V problem} is convex with respect to $\Vb$ and thus can be readily solved by off-the-shelf convex solvers, such as CVX. However, the optimization of the pinching-antenna positions problem in \eqref{eqn: x problem} is challenging. To solve the overall problem, we need to devise an algorithm to solve problem \eqref{eqn: x problem}.

\subsubsection{Solving Problem \eqref{eqn: x problem}}
First, let's recall the MSE of user $m$ as 
\begin{align} \label{eqn: mse reformulation}
    e_m &= \sum_{i=1}^M |u_m \hb_m^\top\vb_i|^2 - 2 \mathcal{R} (u_m\hb_m^\top\vb_m) + \sigma_m^2 |u_m|^2 +1\notag\\
    &= \sum_{i=1}^M \Big|u_m \sum_{n=1}^N h_{m,n} v_{i,n}\Big|^2 - 2 \mathcal{R} \Big\{u_m\sum_{n=1}^N h_{m,n} v_{m,n}\Big\} \notag\\
    & \ \ \ \  +\sigma_m^2 |u_m|^2 +1,
\end{align}
where $(\cdot)^*$ and $\mathcal{R}\{\cdot\}$ return the conjugate and the real part of a complex-valued number, respectively.
With \eqref{eqn: mse reformulation}, the pinching-antenna position optimization in problem \eqref{eqn: x problem} can be equivalently rewritten as
\begin{subequations} \label{p: location problem}
    \begin{align}
        \min_{\tilde \xb} ~& \sum_{m=1}^M w_m^{t+1} \sum_{i=1}^M \Big|u_m^{t+1} \sum_{n=1}^N h_{m,n} v_{i,n}^{t+1}\Big|^2 \notag\\
        & \ \ \ \ \ - 2 w_m^{t+1} \mathcal{R} \Big\{u_m^{t+1}\sum_{n=1}^N h_{m,n} v_{m,n}^{t+1}\Big\} \\
        \st ~& x_{\min} \leq \tilde x_n \leq x_{\max}, \forall n.
    \end{align}
\end{subequations}
Note that problem \eqref{p: location problem} is decomposable with respect to the pinching antennas, which allows us to use the BCD method to sequentially update the positions of the antennas in an iterative manner. 
In particular, the position of the $n$-th pinching antenna is updated by solving the following problem 
\begin{subequations} \label{p: location subproblem}
    \begin{align}
        \min_{\tilde x_n} ~& \sum_{m=1}^M w_m^{t+1} \sum_{i=1}^M \Big|u_m^{t+1} v_{i,n}^{t+1} h_{m,n} + \beta_{m,i,-n}^{t+1}\Big|^2 \notag\\
        &\ \ \ \ - 2 w_m^{t+1} \mathcal{R} \Big\{u_m^{t+1} v_{m,n}^{t+1} h_{m,n} + \gamma_{m,-n}^{t+1}\Big\} \\
        \st ~& x_{\min} \leq \tilde x_n \leq x_{\max}, 
    \end{align}
\end{subequations}
where $\beta_{m,i,-n}^{t+1}$ and $\gamma_{m,-n}^{t+1}$ are defined as
\begin{align*}
    \beta_{m,i,-n}^{t+1} &= u_m^{t+1} \sum_{j\neq n}^N h_{m,j} v_{i,j}^{t+1},\\
    \gamma_{m,-n}^{t+1} &= u_m^{t+1} \sum_{j\neq n}^N h_{m,j} v_{m,j}^{t+1}.
\end{align*}
By inserting \eqref{eqn: channel mu} into \eqref{p: location subproblem}, one can further equivalently rewrite it as \eqref{p: location subproblem 2},
\begin{figure*} 
\begin{subequations} \label{p: location subproblem 2}
    \begin{align}
        \min_{\tilde x_n} ~& \sum_{m=1}^M  \frac{w_m^{t+1} \Big(\sum_{i=1}^M \big|u_m^{t+1} v_{i,n}^{t+1} \big|^2 \Big) \eta}{\big[(\tilde x_n - x_m)^2 + C_{m,n}\big] e^{2 \alpha \tilde x_n}} + 2 w_m^{t+1} \mathcal{R} \left\{ \frac{u_m^{t+1} \Big(\sum\limits_{i=1}^M v_{i,n}^{t+1} \beta_{m,i,-n}^{t+1} -   v_{m,n}^{t+1} \Big) \eta^{\frac{1}{2}} e^{j \theta_{m,n}(\tilde x_n)} }{\big[(\tilde x_n - x_m)^2 + C_{m,n}\big]^{\frac{1}{2}} e^{\alpha \tilde x_n}} - \gamma_{m,-n}^{t+1}\right\} \label{eqn: obj 1}\\
        \st ~& 0 \leq \tilde x_n \leq x_{\max},
    \end{align}
\end{subequations}
\rule{\textwidth}{0.4pt}
\end{figure*}
where $\theta_{m,n}(\tilde x_n) = \frac{2\pi}{\lambda} \big[(\tilde x_n - x_m)^2 + C_{m,n}\big]^{\frac{1}{2}} + \frac{2\pi}{\lambda_g} \tilde x_n$.
As can be seen, the positions of the pinching antennas are complicatedly coupled in the cumulated phase shift, free-space path loss, and in-waveguide attenuation, making problem \eqref{p: location subproblem 2} challenging to solve. Moreover, as $\tilde x_n$ varies along the waveguide, $\theta_{m,n}(\tilde x_n)$ varies rapidly due to the dependence on both the free-space and guided-wave phase shifts. Consequently, the real part term in \eqref{eqn: obj 1} fluctuates considerably, causing the objective function to exhibit severe oscillations. 
This behavior is numerically illustrated in Fig. \ref{fig: mse obj value} for $N = 8$, $M = 8$, $\alpha = 0.0092$ m$^{-1}$, $d_v = 3$ m, $D = 5$ m, $f_c = 6$ GHz, $P_{\max} = 40$ dBm and $\sigma_w^2 = -70$ dBm. 
Due to this highly nonconvex behavior, standard optimization methods (e.g., gradient projection) may converge to undesired local minima of \eqref{eqn: obj 1}.  
\begin{figure}[!t]
	\centering
	\includegraphics[width=0.86\linewidth]{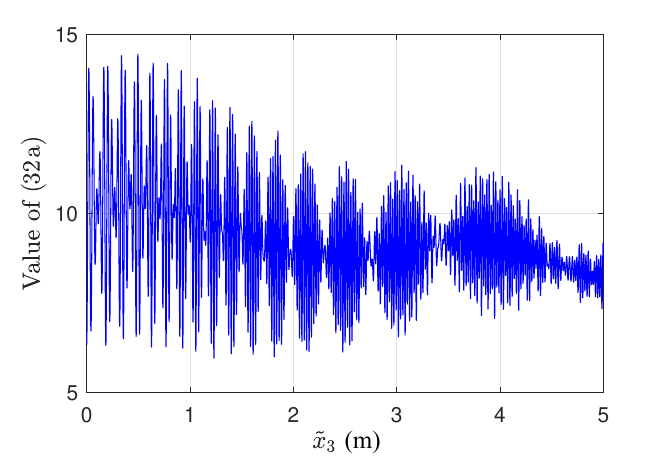}\\
        \captionsetup{justification=justified, singlelinecheck=false, font=small}	
        \caption{The value of \eqref{eqn: obj 1} as $\tilde x_3$ varies along the waveguide.} \label{fig: mse obj value} 
\end{figure} 

On the other hand, observing that \eqref{eqn: obj 1} is just a function of a single real variable $\tilde x_n \in [0,x_{\max}]$, a linear search can be adopted to solve problem \eqref{p: location subproblem 2}. In particular, the linear search approach discretizes the feasible region $[0,x_{\max}]$ and exhaustively evaluates the objective function at each candidate point. 
Following the above steps, the WMMSE-based algorithm iteratively updates each block of variables until converge.


\subsection{Low-Complexity Scheme via MRC-Based Approximation}
From the analysis in Section \ref{subsec: wmmse alg for joint design}, the primary challenges in solving problem \eqref{p: sum rate maximization} stem from two aspects. First, the joint design of the beamformers and pinching-antenna positions necessitates an iterative algorithm that alternates between updating each variable block. Second, the position of each pinching antenna intricately affects the accumulated phase shift, free-space path loss, and in-waveguide attenuation, causing the objective function to exhibit pronounced oscillations. In practice, this necessitates a linear search to determine suitable pinching-antenna positions along the waveguide. As a result, for systems with long waveguides or a high operating frequency, the linear search can become computationally prohibitive, thus limiting its efficiency in practical applications.

In this subsection, we aim to devise a low-complexity algorithm for problem \eqref{p: sum rate maximization} that balances system performance against computational overhead. To this end, we propose a simplified scheme based on MRC beamforming, which iteratively solves two subproblems. Specifically, the first subproblem fixes the beamformer directions according to the MRC principle and optimizes the pinching-antenna positions. Then, the second subproblem optimizes the power allocation coefficients given the obtained pinching-antenna positions. 
This simplified approach, together with a phase-removal approximation to mitigate the severe oscillations, enables a low-complexity gradient projection-based algorithm for the optimization of the pinching-antenna positions, thus significantly reducing algorithm complexity. 
The proposed scheme is detailed below.

\subsubsection{Reformulation of Problem \eqref{p: sum rate maximization} With MRC Beamforming}
By using MRC beamforming, the beamfomer for user $m$ is fixed to the direction of its channel. In particular, beamformer $\vb_m$ is given by 
\begin{align}
    \vb_{m} = \sqrt{\kappa_{m}} \hb^*_{m},
\end{align}
where $\hb^*_{m}$ is the conjugate of $\hb_m$, and $\kappa_{m} \geq 0 $ is the power allocation coefficient for user $m$. Then, the sum rate maximization problem \eqref{p: sum rate maximization} reduces to
\begin{subequations} \label{p: mrc sum rate maximization}
    \begin{align}
        \max_{\tilde \xb, \boldsymbol{\kappa}} &\sum_{m=1}^M \log_2\left(1 + \frac{ \kappa_m \|\hb_m \|^4}{\sum_{i \neq m} \kappa_i |\hb_m^\top \hb_i^*|^2 + \sigma_m^2}\right) \label{eqn: mrc obj 1}\\
        \st~ & \sum_{m=1}^M \kappa_{m} \|\hb_m\|^2  \leq P_{\max}, \label{eqn: mrc power}\\
        & 0 \leq \tilde x_n \leq x_{\max}, \forall n \in \Nset, \label{eqn: mrc antenna location}
    \end{align}
\end{subequations}
where $\boldsymbol{\kappa} = [\kappa_1,...,\kappa_M]^\top$. 
Problem \eqref{p: mrc sum rate maximization} remains challenging due to the nonconvexity of the objective function and its intricate dependence on the pinching-antenna positions. Additionally, the coupling between the power allocation coefficients and antenna positions in the transmit power constraint further complicates optimization.

To address these challenges while maintaining low computational complexity, we adopt a two-stage algorithm to solve problem \eqref{p: mrc sum rate maximization}. In the first stage, we optimize the pinching-antenna positions to maximize the sum rate in \eqref{eqn: mrc obj 1}, assuming fixed power allocation coefficients $\kappa_m$. In the second stage, we optimize the power allocation coefficients to meet the transmit power constraint in \eqref{eqn: mrc power}.

\subsubsection{Pinching-Antenna Position Optimization}
By temporarily omitting the transmit power constraint, we can express the pinching-antenna position optimization subproblem as
\begin{subequations} \label{p: mrc antenna location}
    \begin{align}
        \max_{\tilde \xb} &\sum_{m=1}^M \log_2\left(1 + \frac{ \kappa_m \|\hb_m \|^4}{\sum_{i \neq m} \kappa_i |\hb_m^\top \hb_i^*|^2 + \sigma_m^2}\right) \label{eqn: mrc obj}\\
        \st~ & 0 \leq \tilde x_n \leq x_{\max}, \forall n \in \Nset. 
    \end{align}
\end{subequations}
The main challenge to solve problem \eqref{p: mrc antenna location} arise from the cross terms, $\hb_m^\top \hb_i^*$, in \eqref{eqn: mrc obj}. As the change of the pinching-antenna positions $\tilde x_n$, the value of $|\hb_m^\top \hb_i^*|^2$ oscillates significantly. To mitigate this issue, we adopt an approximation of $|\hb_m^\top \hb_i^*|^2$ using the Cauchy–Schwarz inequality, namely, $|\hb_m^\top \hb_i^*|^2 \leq \|\hb_m\|^2 \|\hb_i^*\|^2$, which effectively removes the rapidly varying phase terms and leads to a more stable and tractable optimization objective. Then, the approximated problem can be written as 
\begin{figure*}
  \begin{subequations} \label{p: mrc antenna location approx}
    \begin{align}
        \max_{\tilde \xb} &\sum_{m=1}^M \log_2\left(1 + \frac{ \kappa_m \bigg(\sum\limits_{n=1}^N \frac{\eta}{ \big[(\tilde x_n - x_m)^2 + C_m\big]e^{2\alpha\tilde x_n}}\bigg)^2}{\sum_{i \neq m} \kappa_i \bigg(\sum\limits_{n=1}^N \frac{\eta}{ \big[(\tilde x_n - x_m)^2 + C_m\big]e^{2\alpha\tilde x_n}}\bigg) \bigg(\sum\limits_{n=1}^N \frac{\eta}{ \big[(\tilde x_n - x_i)^2 + C_i\big]e^{2\alpha\tilde x_n}}\bigg) + \sigma_m^2}\right) \label{eqn: mrc obj approx}\\
        \st~ & 0 \leq \tilde x_n \leq x_{\max}, \forall n \in \Nset. 
    \end{align}
  \end{subequations}  
\rule{\textwidth}{0.4pt}
\end{figure*}
Notably, the objective in \eqref{p: mrc antenna location approx} is decomposable with respect to $\tilde x_n$ which facilitates its solution by the BCD method. To illustrate this, we define
\begin{align}
    A_{m,-n} = \sum\limits_{j\neq n} \frac{\eta}{ \big[(\tilde x_j - x_m)^2 + C_m\big]e^{2\alpha\tilde x_n}}, \forall m \in \Mset,
\end{align} 
Then the subproblem for updating $\tilde x_n$ can be expressed as.
\begin{figure*} 
    \begin{align}\label{p: mrc antenna location approx xn}
        \max_{\tilde x_n \in [0,x_{\max}]} &\sum_{m=1}^M \log_2\left(1 + \frac{ \kappa_m \bigg(\frac{\eta}{ \big[(\tilde x_n - x_m)^2 + C_m\big]e^{2\alpha\tilde x_n}} + A_{m,-n}\bigg)^2}{\sum_{i \neq m} \kappa_i \bigg(\frac{\eta}{ \big[(\tilde x_n - x_m)^2 + C_m\big]e^{2\alpha\tilde x_n}} + A_{m,-n}\bigg) \bigg(\frac{\eta}{ \big[(\tilde x_n - x_i)^2 + C_i\big]e^{2\alpha\tilde x_n}} + A_{i,-n}\bigg) + \sigma_m^2}\right)
    \end{align}
\rule{\textwidth}{0.4pt}
\end{figure*}

Although problem \eqref{p: mrc antenna location approx xn} is still nonconvex, its objective function is significantly simpler than \eqref{eqn: mrc obj} as the phase terms are removed, enabling the use of gradient projection. In particular, by applying gradient projection, we iteratively execute the following two steps until convergence:
\begin{enumerate}
\item[{i)}] Compute the derivative $g'(\tilde{x}_n)$ of \eqref{p: mrc antenna location approx xn} with respect to $\tilde{x}_n$;
\item[{ii)}] Update $\tilde{x}_n \leftarrow \mathrm{Proj}_{[0,x_{\max}]} \{\tilde{x}_n + \tau\,g'(\tilde{x}_n)\}$, where the step size $\tau$ is set by the backtracking line search.
\end{enumerate}
This procedure ensures each update remains feasible and provides a stable, low‐complexity method of optimizing the antenna positions within the BCD framework.

\subsubsection{Power Allocation Optimization}
After obtaining the pinching-antenna positions, we optimize the power allocation variables $\{\kappa_m\}_{m=1}^M$ to maximize the sum rate. The associated problem is given by
\begin{subequations}\label{p: mrc sum rate maximization in text}
\begin{align}
    \max_{\{\kappa_m\}_{m=1}^M} 
    & \quad \sum_{m=1}^M \log_2\biggl(
        1 + \frac{\kappa_m \,\|\mathbf{h}_m\|^4}{\sum_{i \neq m} \kappa_i\,|\mathbf{h}_m^\top \mathbf{h}_i^*|^2 + \sigma_m^2}
    \biggr)
    \label{eqn: mrc obj in text}\\
    \st
    & \quad \sum_{m=1}^M \kappa_m \,\|\mathbf{h}_m\|^2 \;\le\; P_{\max}. \label{eqn: mrc power in text}
\end{align}
\end{subequations}
Similar to problem \eqref{p: mrc antenna location approx xn}, the nonconvex problem \eqref{p: mrc sum rate maximization in text} can be tackled using the gradient projection approach.

First, let's respectively define the objective function and the feasible set as follows: 
\begin{align}
    f(\boldsymbol{\kappa}) &= 
    \sum_{m=1}^M  \log_2\biggl(1 + \frac{\kappa_m \,\|\mathbf{h}_m\|^4}{\sum_{i \neq m} \kappa_i\,|\mathbf{h}_m^\top \mathbf{h}_i^*|^2 + \sigma_m^2} \biggr),\\
    \mathcal{K} &\triangleq 
    \Bigl\{\boldsymbol{\kappa}\,\Big|\,
    \sum_{m=1}^M \kappa_m \|\mathbf{h}_m\|^2 = P_{\max},\;
    \kappa_m \ge 0,\;\forall m
    \Bigr\}.
\end{align}
Then, the gradient projection method iteratively execute the following two steps until convergence:
\begin{itemize}
    \item[{i)}] Compute the gradient vector:
    \begin{align}
        \nabla f(\boldsymbol{\kappa}) =\Bigl[\frac{\partial f}{\partial \kappa_1},\ldots,\frac{\partial f}{\partial \kappa_M}\Bigr]^\top,
    \end{align}
    where $\frac{\partial f}{\partial \kappa_m}$ denotes the partial derivative of $f(\boldsymbol{\kappa})$ with respect to $\kappa_m$.
    \item[{ii)}] Update $\boldsymbol{\kappa} \leftarrow \mathrm{Proj}_{\mathcal{K}} \{\boldsymbol{\kappa} + \xi \nabla f(\boldsymbol{\kappa})\}$, where the step size $\xi$ is set by the backtracking line search. 
\end{itemize}
The proposed MRC-based scheme iteratively solves the pinching-antenna position optimization problem \eqref{p: mrc antenna location approx xn} and the power allocation problem \eqref{p: mrc sum rate maximization in text} until some convergence criterion is satisfied. The details are summarized in Algorithm \ref{alg:mrc_based}.

\begin{algorithm}[t] \small
\caption{MRC Beamforming-based Low-Complexity Pinching-Antenna Position Optimization}
\label{alg:mrc_based}
\begin{algorithmic}[1]
\STATE \textbf{Input:} User locations $\{\psib_m\}$, system parameters, initial pinching-antenna positions $\tilde \xb^{(0)}$, fixed MRC beamforming directions.
\STATE \textbf{Initialization:} Equal power allocation $\kappa_m^{(0)}$, set iteration counter $t \leftarrow 0$.

\REPEAT
    \STATE \textbf{(Pinching-Antenna Position Update)}
    \FOR{each antenna $n = 1,\dots,N$}
        \STATE \quad Fix $\{\tilde x_j\}_{j\neq n}$ and $\boldsymbol{\kappa}^{(t)}$;
        \STATE \quad Solve subproblem~\eqref{p: mrc antenna location approx xn} via gradient projection:
        \begin{itemize}
            \item Compute gradient $g'(\tilde x_n)$.
            \item Update $\tilde{x}_n^{(t+1)} \leftarrow \mathrm{Proj}_{[0,x_{\max}]}\{\tilde{x}_n^{(t)} + \tau g'(\tilde{x}_n^{(t)})\}$ with backtracking line search.
        \end{itemize}
    \ENDFOR

    \STATE \textbf{(Power Allocation Update)}
    \STATE \quad With the updated $\tilde \xb^{(t+1)}$, solve the power allocation problem~\eqref{p: mrc sum rate maximization in text} via gradient projection:
    \begin{itemize}
        \item Compute gradient $\nabla f(\boldsymbol{\kappa})$.
        \item Update $\boldsymbol{\kappa}^{(t+1)} \leftarrow \mathrm{Proj}_{\mathcal{K}}\{\boldsymbol{\kappa}^{(t)} + \xi \nabla f(\boldsymbol{\kappa}^{(t)})\}$ with backtracking line search.
    \end{itemize}
    
    \STATE $t \leftarrow t+1$
\UNTIL{Convergence criterion is satisfied}
\STATE \textbf{Output:} Optimized pinching-antenna positions $\tilde \xb^\star$ and power allocations $\boldsymbol{\kappa}^\star$.
\end{algorithmic}
\end{algorithm}

Based on the MRC-based algorithm, we can devise a low-complexity two-stage algorithm.
Specifically, in the first stage, the MRC-based approach is used to first optimize the pinching-antenna positions. Then, in the second stage, given the obtained pinching-antenna positions, the conventional WMMSE algorithm is applied to optimize the beamforming matrix. 
The details of this two-stage algorithm is summarized in Algorithm \ref{alg: wmmse-mrc}.
This two-stage algorithm avoids the need to perform a linear search over the pinching-antenna positions in every iteration of the WMMSE algorithm proposed in Subsection \eqref{subsec: wmmse alg for joint design}, thereby significantly reducing the overall computational complexity.

\begin{algorithm}[t] \small
\caption{Two-Stage WMMSE-MRC Algorithm for Pinching-Antenna Position and Beamformer Optimization} \label{alg: wmmse-mrc}
\begin{algorithmic}[1]
\STATE \textbf{Input:} User locations $\{\psib_m\}$ and system parameters.

\STATE \textbf{Stage 1}:  Run Algorithm \ref{alg:mrc_based} to obtain the pinching-antenna positions, $\tilde x_n^*, \forall n \in \Nset$.           

\STATE \textbf{Stage 2:  WMMSE-based Beamformer Design}
\STATE Update the channel using $\tilde x_n^*, \forall n \in \Nset$.
\STATE Initialize the beamformer using the MRC-based beamformer
\REPEAT  
    \STATE \textbf{Receiver update}: Update $\{u_m\}$ by \eqref{eqn: update of u}.   
    \STATE \textbf{Weight update}:   Update $\{w_m\}$ by \eqref{eqn: update of w}.   
    \STATE \textbf{Beamformer update}: Update the beamformer $\mathbf V$ by solving convex problem \eqref{eqn: V problem}. 
\UNTIL{Convergence criterion is satisfied}

\STATE \textbf{Output:} Final pinching-antenna positions, $\tilde x_n^*, \forall n \in \Nset$ and beamformers $\mathbf V^{*}$.
\end{algorithmic}
\end{algorithm}

\section{Numerical Simulations}  \label{sec: simulation}
In this section, we evaluate the performance of the considered pinching-antenna systems and the proposed algorithms via computer simulations. The simulation parameters are chosen as follows. The in-waveguide attenuation coefficient is set as $\alpha = 0.08$ dB/m \cite{bauters2011planar}, the noise power is set to $-70$ dBm, $f_c = 28$ GHz, $d_v = 3$ m, and $n_{\rm neff} = 1.4$. The choices for the total transmit power $P_{\max}$, the side length $D$ of the square area, the number of users $M$, and the number of pinching antennas $N$ are specified in each figure. In the simulations, the results are obtained by averaging over $20$ random user position realizations. The stopping condition for all iteration-based schemes in this work is that the absolute value of the difference in the objective function between two successive iterations becomes less than $10^{-4}$ or the maximum iteration number $T_{\max} = 20$ is reached.

\subsection{Pinching-Antenna System Versus Conventional Antenna System}
In this subsection, we evaluate the achievable data rate of the pinching-antenna systems, where the conventional fixed-position antenna system is used as a benchmark,  in the presence of in-waveguide attenuation. For fairness, the BS of the reference system employs a uniform linear array using the same number of antennas, i.e., $M$. The center of the array is placed at the central of the communication area, i.e., $[D/2,0,d_v]$, and the inter‑antenna spacing is the classical half wavelength, $\Delta = \frac{\lambda}{2}$.  Because the positions of the reference antennas are fixed, only the digital beamformers need to be optimized. We solve the corresponding sum‑rate maximization problem with the standard WMMSE algorithm.

\begin{figure}[!t]
	\centering
	\includegraphics[width=0.96\linewidth]{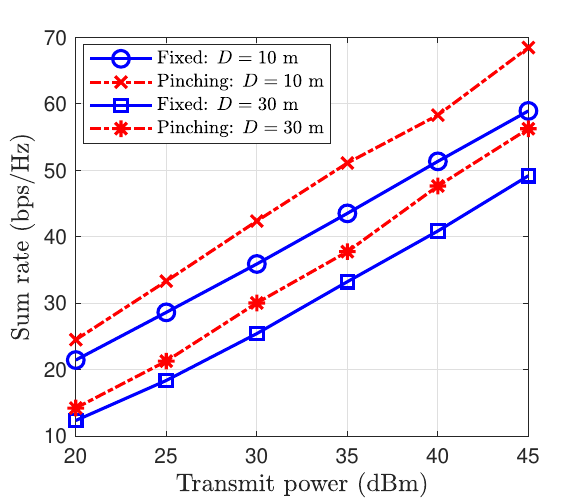}\\
        \captionsetup{justification=justified, singlelinecheck=false, font=small}	
        \caption{Achievable sum rate versus total transmit power $P_{\max}$ with $M=N=8$.} \label{fig: fix rate power}  
\end{figure} 

\begin{figure}[!t]
	\centering
	\includegraphics[width=0.96\linewidth]{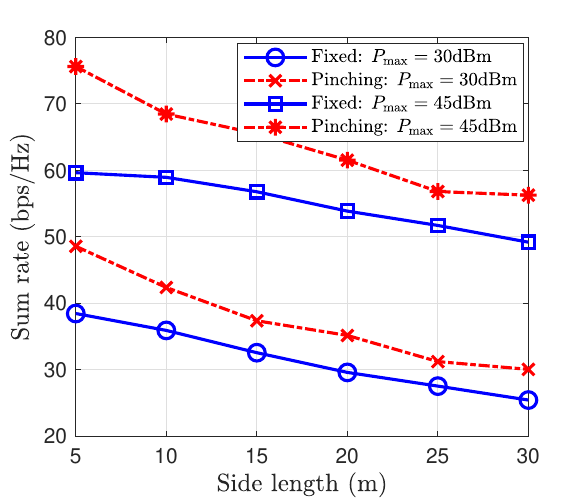}\\
        \captionsetup{justification=justified, singlelinecheck=false, font=small}	
        \caption{Achievable sum rate versus side length $D$ with $M=N=8$.} \label{fig: fix rate d} 
\end{figure}

In Fig. \ref{fig: fix rate power}, we compare the achievable sum rate of the pinching‑antenna system to that of the conventional fixed-position antenna system for two coverage‑area side‑lengths, i.e., $D \in \{10,30\}$ m. 
As can be observed from this figure, across all transmit power levels, the pinching-antenna system (red dashed lines) consistently outperforms the fixed-position antenna system (blue solid lines). The superior performance is due to the ability of pinching antennas to strengthen LoS links while reducing inter‑user interference. 
On the other hand, enlarging the service area from $10$ m to $30$ m reduces the data rates of both systems because users are, on average, farther away from the antennas. 
This advantage of the pinching-antenna system with different communication area side length is further validated by Fig. \ref{fig: fix rate d}. As seen, the pinching-antenna system consistently outperforms the conventional fixed-antenna system over all considered communication region sides.

\subsection{Performance of the Proposed Algorithms}

\begin{figure}[!t]
	\centering
	\includegraphics[width=0.96\linewidth]{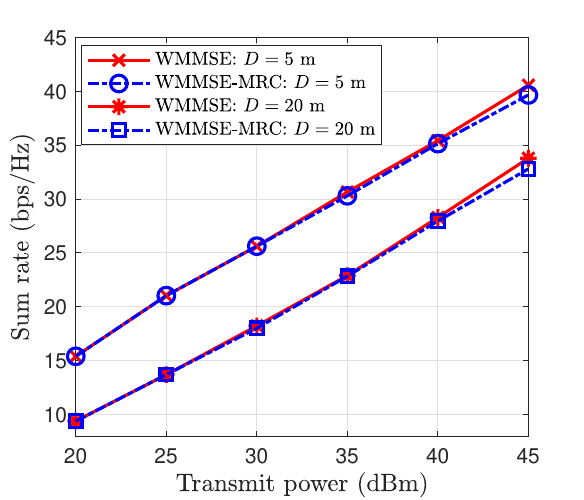}\\
        \captionsetup{justification=justified, singlelinecheck=false, font=small}	
        \caption{Achievable sum rate comparison for the proposed WMMSE-based algorithm and the proposed two-stage WMMSE-MRC algorithm with $M=N=4$.} \label{fig: wmmse-mrc power} 
\end{figure} 

In this subsection, we examine the performance of the proposed optimization algorithms for pinching-antenna systems in terms of the achievable data rate and the computational complexity.
Firstly, Fig. \ref{fig: wmmse-mrc power} compares the achievable sum rate of the proposed WMMSE-based algorithm with the proposed two-stage WMMSE-MRC algorithm for different transmit power levels and coverage area sizes. 
Specifically, the step size for the linear search for the pinching-antenna positions in the WMMSE method is set as $\delta = \frac{\lambda_g}{50}$.
As can be observed, the WMMSE-MRC scheme achieves sum rates very close to those of the full WMMSE algorithm for all considered settings. This demonstrates that the MRC-based initialization offers a low-complexity antenna placement strategy that balances the in-waveguide loss and the free-space loss. The WMMSE-MRC method does not jointly optimize the antenna positions and beamformers, and thus avoids the iterative and computationally intensive position search in each WMMSE iteration, which significantly reduces complexity while incurring only a negligible performance loss. The rate gap remains small even for the larger coverage areas ($D = 20$ m), validating the robustness of the two-stage approach.

\begin{figure}[!t]
	\centering
	\includegraphics[width=0.96\linewidth]{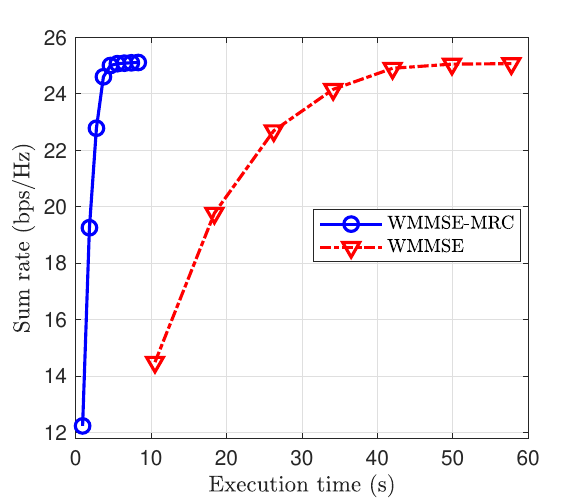}\\
        \captionsetup{justification=justified, singlelinecheck=false, font=small}	
        \caption{Execution time comparison of the proposed WMMSE-based algorithm and the two-stage WMMSE-MRC algorithm with $D=5$ m and $M=N=4$.} \label{fig: rate iteration} 
\end{figure}

Fig. \ref{fig: rate iteration} further shows the achievable sum rate of the proposed WMMSE-MRC and WMMSE algorithms versus the execution time. As illustrated, the WMMSE-MRC algorithm achieves near-optimal performance within a few iterations and a much lower execution time. In contrast, the conventional WMMSE method requires significantly longer runtimes to reach comparable rates. The key reason for this disparity lies in the linear search procedure used by the WMMSE method to optimize each pinching-antenna position, which becomes increasingly time-consuming as the number of antennas and the resolution of the search grid grow. In contrast, the WMMSE-MRC algorithm adopts a low-complexity gradient-based update in the first stage, followed by beamformer refinement in the second stage, thus avoiding the costly exhaustive search. This demonstrates that the WMMSE-MRC approach offers a favorable trade-off between performance and complexity, making it better suited for practical systems with stringent computational complexity constraints.

\subsection{Performance of the Pinching-Antenna System for Different System Parameters}

\begin{figure}[!t]
	\centering
	\includegraphics[width=0.96\linewidth]{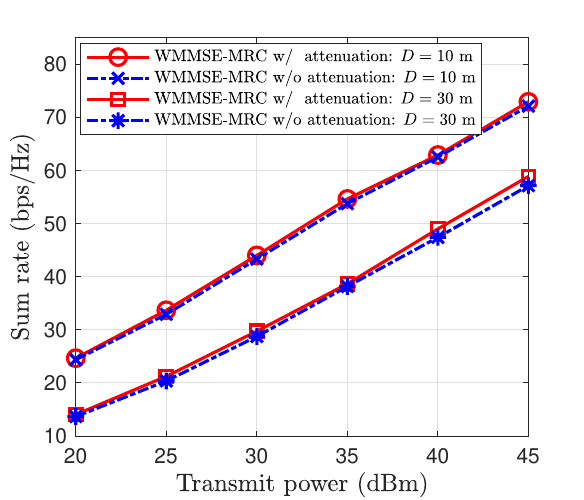}\\
        \captionsetup{justification=justified, singlelinecheck=false, font=small}	
        \caption{Achievable sum rate comparison of the schemes with and without accounting for in-waveguide attenuation with $M=N=8$.} \label{fig: proposed baseline power} 
\end{figure}
\begin{figure}[!t]
	\centering
	\includegraphics[width=0.96\linewidth]{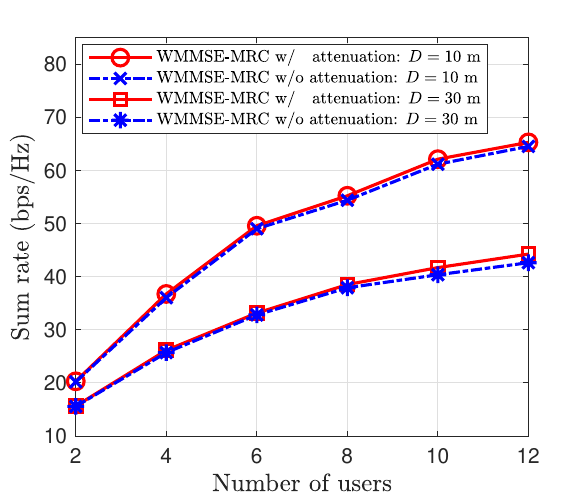}\\
        \captionsetup{justification=justified, singlelinecheck=false, font=small}	
        \caption{Achievable sum rate comparison of the schemes with and without accounting for in-waveguide attenuation versus number of users with $P_{\max} = 35$ dBm and $N = 12$.} \label{fig: proposed baseline user}  
\end{figure}

 
In this subsection, we evaluate the system performance of the pinching-antenna system with difference system parameters.
Fig. \ref{fig: proposed baseline power} compares the achievable sum rate of the schemes with and without accounting for in-waveguide attenuation for two coverage-area side lengths, $D \in \{10, 30\}$ m. The scheme with attention corresponds to the scheme optimizes antenna positions and evaluates performance by fully considering the in-waveguide attenuation. In contrast, the scheme without attention ignores in-waveguide attenuation during the antenna position optimization but includes it in the final sum rate calculation. 
As observed, the scheme with attenuation achieves slightly higher sum rates compared to the scheme without attenuation. This improvement stems from optimizing the antenna positions while accounting for in-waveguide attenuation, which enables a more accurate trade-off between minimizing in-waveguide loss and reducing free-space path loss. 
Meanwhile, Fig. \ref{fig: proposed baseline user} shows the achievable sum rate of the two schemes versus the number of users. As expected, the sum rate increases with the number of users due to enhanced spatial multiplexing, and the proposed scheme consistently outperforms the baseline in all considered cases.

\begin{figure}[!t]
	\centering
	\includegraphics[width=0.96\linewidth]{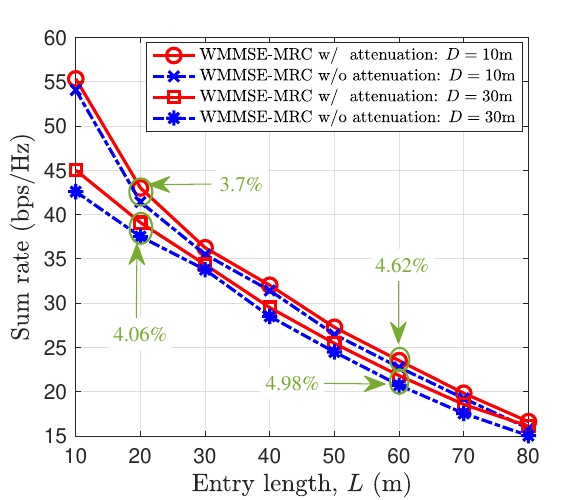}\\
        \captionsetup{justification=justified, singlelinecheck=false, font=small}	
        \caption{Achievable sum rate comparison of the schemes with and without accounting for in-waveguide attenuation versus entry length with $P_{\max} = 45$ dBm and $M= N = 8$.} \label{fig: rate L} 
\end{figure} 

To evaluate the performance of the proposed algorithm in diverse application scenarios, we consider a system where the feed point is deployed outside the communication area. The total waveguide length is $L+D$, with the segment from $0$ to $L$ routed outside the user region, and users distributed only within the interval $[L, L+D]$. This configuration reflects practical deployments where pinching antennas are routed from an outdoor feed point to serve multiple indoor users, such as in large buildings, shopping centers, or smart industrial sites.
As shown in Fig.~\ref{fig: rate L}, the overall sum rate decreases as the entry length $L$ increases, due to the higher propagation loss along the waveguide. More importantly, the relative rate loss exhibits an upward trend with both increasing entry length $L$ and communication region size $D$. Here, the relative rate loss is defined as $(R_{\rm w.~att.} - R_{\rm w.o. ~att.})/R_{\rm w.~att.} \times 100\%$, where $R_{\rm w.~att.}$ and $R_{\rm w.o.~att.}$ denote the sum rates achieved with and without accounting for in-waveguide attenuation during system design, respectively.
This observation suggests that the impact of in-waveguide attenuation becomes more significant when either the entry length $L$ or the communication region size $D$ increase. Nevertheless, the proposed algorithm demonstrates satisfactory system performance even in expansive deployment scenarios, highlighting its robustness and practical applicability in diverse communication scenarios.

\section{Conclusion} \label{sec: conclusion}
This paper investigated the downlink design of pinching-antenna systems under practical considerations by explicitly accounting for in-waveguide attenuation.
We began with a single-user scenario and derived the closed-form globally optimal antenna placement. Building on this, we conducted a detailed theoretical analysis of the achievable rate loss caused by neglecting in-waveguide attenuation. Closed-form expressions and tight performance bounds were established, offering quantitative design rules that directly relate system parameters—such as the attenuation level, waveguide height, and coverage area—to the expected performance loss.
These analytical results provide not only crucial insights but also practical guidelines for system design, showing that in-waveguide attenuation can be safely ignored under mild conditions.
To demonstrate broader applicability, we extended the study to the MU-MIMO case and proposed efficient algorithmic solutions based on WMMSE and MRC approximations.
Extensive simulations validated the theoretical findings and showed that pinching-antenna systems significantly outperform conventional fixed-antenna designs, confirming their potential as a flexible and high-performance architecture for future wireless networks.

\begin{appendices}
    \section{Proof of Proposition \ref{prop: expected rate loss}} \label{appd: expected rate loss} 
    We first focus on the scheme without attenuation, i.e., Scheme $1$. The associated instantaneous received SNR is given by
    \begin{align} \label{eqn: snr wo}
        \SNR_{\wo} = \frac{\rho \eta}{C e^{2\alpha \bar x}}.
    \end{align}

    For the scheme with attenuation, i.e., Scheme $3$, the associated SNR is presented in the following lemma.
    \begin{lem} \label{lem: snr w attenuation} 
        The instantaneous received SNR for the scheme with attenuation can be approximately written as
        \begin{align} \label{eqn: snr w}
            \SNR_{\w} \approx \frac{\rho \eta}{C (1 - \alpha^2 C) e^{2\alpha \bar x}}
        \end{align}
    \end{lem}
    \begin{IEEEproof}
        See Appendix \ref{appd: snr w attenuation}.
    \end{IEEEproof}

    Based on \eqref{eqn: snr wo} and \eqref{eqn: snr w}, one can see that $\SNR_{\wo} = \SNR_{\w}(1 - \alpha^2 C)$. Then, the instantaneous rate loss is given by
    \begin{align}
        \Delta R &= R_{\w} - R_{\wo} \notag\\
        & =\log_2 (1 + \SNR_{\w}) - \log_2 (1 + \SNR_{\wo})\notag\\
        & = \log_2 \left( \frac{1 + \SNR_{\w}}{1 + \SNR_{\w}(1 - \alpha^2 C)}\right)
    \end{align}
    In the high SNR regime, the instantaneous rate loss can be approximated by
    \begin{align}
        \Delta R \approx - \log_2 \left(1 - \alpha^2 (\bar y^2 + d_v^2)\right).
    \end{align}
    Then, the averaged rate loss is given by
    \begin{align}
        \mathbb{E}[\Delta R] &= - \frac{2}{D}\int_{0}^{\frac{D}{2}} \log_2 \left( 1 -  \alpha^2 (\bar y^2 + d_v^2) \right) d \bar y \notag\\
        &\overset{(a)}{\approx} \frac{2}{D}\int_{0}^{\frac{D}{2}} \frac{\alpha^2 (\bar y^2 + d_v^2)}{\ln 2} d \bar y \notag\\
        & = \frac{\alpha^2}{\ln 2} \left(\frac{D^2}{12} + d_v^2 \right),
    \end{align}
    where, in (a), the first-order Taylor approximation is used. This completes the proof.
    \hfill $\blacksquare$

    \section{Proof of Lemma \ref{lem: snr w attenuation}} \label{appd: snr w attenuation} 
    First, recall that the pinching-antenna position for Scheme $3$ is $\tilde x^* = \bar{x} + \frac{-1 + \sqrt{1 - 4\alpha^2 C}}{2\alpha}$. Noting that the waveguide attenuation $\alpha$ is generally a small number, the term $\sqrt{1 - 4\alpha^2 C}$ can be approximated by $1 - 2\alpha^2 C$. Therefore, the pinching-antenna position can be approximately represented by
    \begin{align}
        \tilde x^* \approx \bar x - \alpha C.
    \end{align}
    With this antenna position, the received SNR can be approximately written as
    \begin{align}
        \SNR_{\w} &\approx \frac{\rho \eta}{\alpha^2 C^2 e^{2 \alpha (\bar x - \alpha C)} + C e^{2\alpha(\bar x - \alpha C)}} \notag\\
        & =  \frac{\rho \eta}{ C (1+\alpha^2 C) e^{2 \alpha \bar x} e^{-2\alpha^2 C}} \notag\\
        & \overset{(a)}{\leq} \frac{\rho \eta}{C (1 - \alpha^2 C)e^{2 \alpha \bar x}}
    \end{align}
    where step (a) is obtained due to the fact that $ e^{2\alpha^2 C} \leq \frac{1 + \alpha^2 C}{1 - \alpha^2 C}$, if $0 \leq \alpha^2 C < 1$ \cite{abramowitz1965handbook}. 
    This completes the proof. 

    \hfill $\blacksquare$
\end{appendices}


\smaller[1]

\end{document}